\documentclass[twocolumn]{aastex61}
\pdfoutput=1

\usepackage{tcolorbox}
\usepackage{savesym}
\usepackage{soul}

\usepackage{amsmath}
\savesymbol{tablenum}
\usepackage{verbatim}
\usepackage{siunitx}
\restoresymbol{SIX}{tablenum}
\definecolor{orcidlogocol}{HTML}{A6CE39}
\usepackage{natbib}
\usepackage{hyperref}
\usepackage{graphicx}
\usepackage{float}

\newcommand{\mathbold}[1]{\mbox{\boldmath $\bf#1$}}

\newcommand\mubold{{\mathbold \mu}}

\begin{document}
	
	\title{Revisiting MOA 2013-BLG-220L: A Solar-type star with a cold super-Jupiter companion}

	\correspondingauthor{Aikaterini Vandorou}
	\email{aikaterini.vandorou@utas.edu.au}

		\author{Aikaterini Vandorou}
\affil{School of Natural Sciences, University of Tasmania,
	Private Bag 37 Hobart, Tasmania, 7001, Australia}
	
		\author{David P.~Bennett}
\affil{Laboratory for Exoplanets and Stellar Astrophysics, NASA/Goddard Space Flight Center, Greenbelt, MD 20771, USA}
\affil{Department of Astronomy, University of Maryland, College Park, MD 20742, USA}
	
	\author{Jean-Philippe~Beaulieu}
\affil{School of Natural Sciences, University of Tasmania,
	Private Bag 37 Hobart, Tasmania, 7001, Australia}
\affil{Sorbonne Universit\'es, UPMC Universit\'e Paris 6 et CNRS, 
	UMR 7095, Institut d'Astrophysique de Paris, 98 bis bd Arago,
	75014 Paris, France}

		\author{Christophe~Alard}
	\affil{Sorbonne Universit\'es, UPMC Universit\'e Paris 6 et CNRS, 
		UMR 7095, Institut d'Astrophysique de Paris, 98 bis bd Arago,
		75014 Paris, France}
	
			\author{Joshua~W.~Blackman}
	\affil{School of Natural Sciences, University of Tasmania,
		Private Bag 37 Hobart, Tasmania, 7001, Australia}
	
	\author{Andrew A.~Cole}
	\affil{School of Natural Sciences, University of Tasmania,
		Private Bag 37 Hobart, Tasmania, 7001, Australia}
	
		\author{Aparna~Bhattacharya}
	\affil{Laboratory for Exoplanets and Stellar Astrophysics, NASA/Goddard Space Flight Center, Greenbelt, MD 20771, USA}
	\affil{Department of Astronomy, University of Maryland, College Park, MD 20742, USA}

		\author{Ian~A.~Bond}
	\affil{Institute of Natural and Mathematical Sciences, Massey University, Auckland 0745, New Zealand}

	       \author{Naoki  Koshimoto}
	\affil{Okayama Astrophysical Observatory, National Astronomical Observatory of Japan, Asakuchi, Okayama 719-0232, Japan}
	
		\author{Jean-Baptiste~Marquette}
	\affil{Laboratoire d'astrophysique de Bordeaux, Univ. Bordeaux, CNRS, B18N, allée Geoffroy Saint-Hilaire, 33615 Pessac, France}
        \affil{Sorbonne Universit\'es, UPMC Universit\'e Paris 6 et CNRS, 
		UMR 7095, Institut d'Astrophysique de Paris, 98 bis bd Arago,
		75014 Paris, France}

	\begin{abstract}

We present the analysis of high-resolution images of MOA-2013-BLG-220, taken with the Keck adaptive optics system 6 years after the initial observation, identifying the lens as a solar-type star hosting a super-Jupiter mass planet. The masses of planets and host-stars discovered by microlensing are often not determined from light curve data, while the star-planet mass-ratio and projected separation in units of Einstein ring radius are well measured. High-resolution follow-up observations after the lensing event is complete can resolve the source and lens. This allows direct measurements of flux, and the amplitude and direction of proper motion, giving strong constraints on the system parameters. Due to the high relative proper motion, $\mubold_{\rm rel,Geo} = 12.62\pm0.11$ mas/yr, the source and lens were resolved in 2019, with a separation of $77.1\pm0.5$ mas. Thus, we constrain the lens flux to $K_{\rm Keck,lens}= 17.92\pm0.02$. By combining constraints from the model and Keck flux, we find the lens mass to be $M_L = 0.88\pm0.05\ M_\odot$ at $D_L = 6.72\pm0.59$ kpc. With a mass-ratio of $q=(3.00\pm0.03)\times10^{-3}$ the planet's mass is determined to be $M_P = 2.74\pm0.17\ M_{J}$ at a separation of $r_\perp = 3.03\pm0.27$ AU. The lens mass is much higher than the prediction made by the Bayesian analysis that assumes all stars have an equal probability to host a planet of the measured mass ratio, and suggests that planets with mass ratios of a few 10$^{-3}$ are more common orbiting massive stars. This demonstrates the importance of high-resolution follow-up observations for testing theories like these.

	\end{abstract}


	\section{Introduction} \label{sec:intro}
	
Studying planets that lie beyond the snow-line is key to understanding the core-accretion theory (\cite{Lissauer1993}; \cite{Ida2004}; \cite{Kennedy2006}). This parameter space is not easily probed by detection methods such as radial velocity and stellar transits. Gravitational microlensing is currently the only method with enough sensitivity to detect cold low-mass planets around nearby stars, as well as stars in the Galactic Bulge. This is because the technique is not dependent on the luminosity of the host star.

A  limitation posed by gravitational microlensing, however, is the relatively low precision for physical parameter measurements, which is a consequence of uncertain host-star distance and mass. On the other hand, relative physical parameters, such as the mass-ratio, can be determined accurately for most events from the photometric light curve. 
With the help of high angular resolution follow-up observations taken 5-10 years after peak magnification, the light from the lens and source star can be accurately measured and constraints placed on their physical properties. Furthermore, for some events this time will have allowed the lens and source star to have separated enough to be observed independently \citep{Bennett2007, bennett15, Bhattacharya2018}.

Observations with high angular resolution can be conducted with Keck, VLT, Magellan or Subaru with the adaptive optics system. Space based telescopes, such as the Hubble Space Telescope (HST) can also be used. An example of a follow-up observation where source and lens were resolved with a separation of $\sim 60$ mas, is OGLE 2005-BLG-169 \citep{Batista2015,bennett15}. This event was observed with Keck's NIRC2 adaptive optics system 8 years after the peak magnification occurred. With the new Keck data as well as data obtained from HST, the initial model was refined and the parameters constrained, revealing the system to be a Uranus-mass planet orbiting a K5-type main sequence star at a distance of 4 kpc. 

The light curve modeling of the planetary microlensing event MOA-2013-BLG-220 revealed a large relative proper motion, and therefore a high chance of resolving both the lens and the source star. In this paper we perform a new light curve analysis as well as present high resolution Keck follow-up observations of MOA-2013-BLG-220, in order to constrain the lens flux and the relative source-lens proper motion. Using these new constraints we can revisit the physical parameters of the system.

	\section{Microlensing Event MOA-2013-BLG-220}
	\label{MOA}
Microlensing event MOA-2013-BLG-220 was identified and announced on 1st April 2013 by the Microlensing Observations in Astrophysics (MOA) collaboration \textbf{\citep{Bond2001}}.
An additional alert was issued 46 hours later by $\mu$FUN which stated that the event was likely to be a high magnification event. This resulted in the Optical Gravitational Lens Experiment (OGLE, \textbf{\cite{Udalski1994}, \cite{Udalski2003}}) switching their telescope into ``follow-up'' mode, since it is typically dedicated to survey operations. MOA-2013-BLG-220 actually lay between OGLE's mosaic camera CCD chips, however since this was seen as an `interesting' event, OGLE altered their pointing and increased their cadence. It was also observed by the Las Cumbres Observatory Global Telescope (LCOGT).
 The data, modeling and analysis of the light curve are presented in \cite{Yee2014}, where a planet was discovered with a relatively high mass ratio of $q = (3.01 \pm 0.02) \times 10^{-3}$.  Using the microlensing parameters (Einstein and source crossing time, $t_E$ and $t_*$, respectively) and the angular size of the source,  $\theta_*$, the Einstein ring radius was derived to be $\theta_E = 0.456 \pm 0.073$ mas. 

No Bayesian estimates of the physical parameters of the system were made by \cite{Yee2014}, as the results would have not been informative without microlensing parallax constraints. Constraints from the Einstein-ring radius mass-distance relation, the main-sequence mass-luminosity relation, and the Galactic rotation curve led the authors to conclude that the lens was most likely a low-mass disk star for maximum consistency with the light curve model. Thus, they predict that the lens should be at a distance $D_L < 6.5$ kpc with a mass $M_L < 0.77$ $M_\odot$. This would imply a planetary companion of $M_{p} < 2.4$ $M_J$.

\cite{Yee2014} find a relative proper motion between the source star and lens of $12.5\pm 1.0$ mas/yr. This meant that the event would be a good candidate for high-resolution follow up observations as the source and the lens will have separated within the next ten years since the initial observation. 

\begin{deluxetable}{ c | c c }
\tablecaption{Light Curve Model Parameters  \label{tab-mparams} }
\tablewidth{0pt}
\tablehead{
\colhead{Parameter}  & \colhead{Model} & \colhead{Yee+14} 
}  
\startdata
	$t_E$ (days) & $13.31\pm 0.08$ &$13.23 \pm 0.03$  \\   
	$t_0$ (${\rm HJD}^\prime$) & $6386.9204\pm 0.0012$ & $6386.9199\pm 0.0009$ \\
	$u_0$ &$-0.013142\pm 0.00015$  & $-0.01323 \pm 0.0004$ \\
	$s$ & $0.98576 \pm 0.00012$ & $0.9857 \pm$ 0.0001 \\
     $\alpha$ (rad) & $-1.4206 \pm 0.0045$ & $-1.4224 \pm 0.0030$  \\
	$q \times 10^{3}$ & $2.999 \pm 0.027$ & $3.01 \pm 0.02$  \\
	$t_\ast$ (days) & $0.02042\pm 0.00016$ & 0.02037 \\
	$I_s$ & $19.232 \pm 0.021$ & $19.205 \pm0.003$ \\
	$V_s$ & $20.794 \pm 0.021$ & - \\
	$\theta_E$(mas) & 0.457 $\pm$ 0.005 & 0.456 $\pm$ 0.073 \\ 
	$\chi^2$ & 6039.12 & -\\
	for 6475 dof & & \\
\enddata
\end{deluxetable}

\begin{figure*}
	\centering
	\includegraphics[width=12cm]{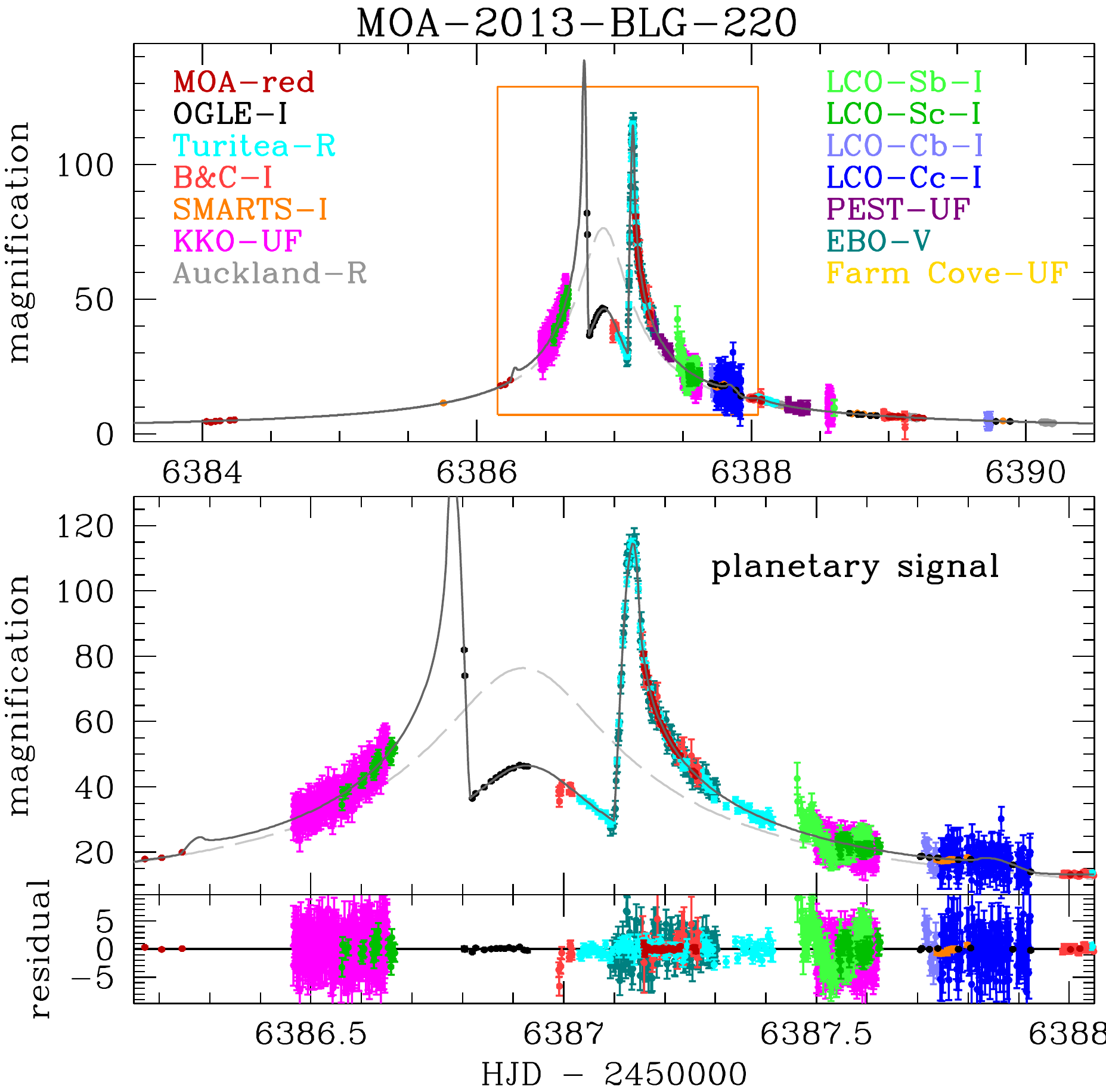}
	\caption{ The MOA-2013-BLG-220 light curve with the best fit model presented in this paper. This
	model differs very slightly from the model presented in the discovery paper \citep{Yee2014}. Data from different observatories are indicated by the different colors.} 
	\label{fig-lc_model}
\end{figure*}

\section{Microlensing Light Curve Model}
\label{sec-lc_model}

It was necessary to perform a new light curve analysis for MOA-2013-BLG-220 in order to obtain a Markov Chain Monte Carlo (MCMC) distribution of parameters. These parameters will be used to examine the effect of
model uncertainties on the uncertainties in the physical parameters of the planetary system that
serves as the lens.

The reanalyses uses {\bf almost} the same image data as \cite{Yee2014}, but the 
MOA, CTIO and B$\&$C data were re-reduced using the procedure outlined in \cite{Bond2017}. 
This includes detrending of the MOA data to remove systematic errors such as chromatic differential 
refraction, which can have a significant effect on microlensing light curves \citep{bennett12}. This detrending
was not included in the original analysis, however it did not have a large effect.
 In addition, the \cite{Yee2014} analysis did not include the Las Cumbres Observatory (LCO) and
South African Astronomical Observatory (SAAO) Telescope-B data. \cite{Yee2014} did not explain
why this data set were excluded.

The modeling used the image-centered ray-shooting method \citep{Bennett1996,bennett-method},
and the best fit light curve is shown in Figure~\ref{fig-lc_model}, with model parameters given in
Table~\ref{tab-mparams}. Our MCMC calculation was used to determine the distribution of light curve
parameters that are consistent with the data. The model parameters given in Table~\ref{tab-mparams}
are the mean and root-mean-square (RMS) from the MCMC calculation. The parameters include three parameters that
also apply to single lens events: the Einstein radius crossing time, $t_E$, the time of closest approach
between the source and lens center-of-mass, $t_0$, and the lens-source separation at this 
closest approach, $u_0$, in units of the Einstein radius. Four more parameters are required to account
for the planet: the planet-star mass ratio, $q$, their separation, $s$ (in units of the Einstein radius), the
angle, $\alpha$ between the planet-star separation vector and the source trajectory, and the source
radius crossing time, $t_*$. Our model parameters are generally quite similar to the results
from \citet{Yee2014}, but there are some differences due to the slightly different data sets. 

The source magnitudes from the MOA-red and CTIO-$V$ passbands were calibrated to the 
OGLE-III catalog \citep{ogle3_catalog} to determine the source magnitudes in the $I$ and $V$ 
passbands, $I_s$ and $V_s$, given in Table~\ref{tab-mparams}. The uncertainties in these magnitudes
include an estimated 0.02 magnitude calibration uncertainty.
The average \textit{I} band magnitude and color are:
  \begin{equation}
    (I, V-I)_{S} = (19.232, 1.562) \pm (0.021, 0.029) \ .
    \label{eq-VI}
  \end{equation}

In order to derive the angular source radius, we determined the centroid of the red clump magnitude for stars within
2 arcmin from the target in the OGLE-III photometry catalog. Then we compared the observed magnitude and color to the intrinsic extinction corrected magnitudes for Galactic bulge red clump stars at the Galactic longitude for this event to determine the extinction \citep{nataf13}. This yields
an extinction of $A_I = 1.175\pm 0.050$ and $E(V-I) = A_I-A_V = 0.970\pm 0.025$, and from this, we find the dereddened source color and magnitude to be $(I, V-I)_{S,0} = (18.057, 0.592) \pm (0.054, 0.038)$.
 Using the surface brightness relationship from \cite{Boyajian2013}: 
 \begin{equation}
  \log(2\theta_{*}) = 0.5014+0.4197(V-I)_{0} -0.2I_{0} \ ,
      \label{eq-thetaStar}
 \end{equation}
 we calculate the source star angular radius to be $\theta_* = 0.689 \pm 0.052\ \rm \mu as$, which is slightly smaller but compatible with the \cite{Yee2014} value of $\theta_{*} = 0.704\ \rm \mu as$. 
 
\section{Follow-up Observations with KECK}
A follow-up observation of MOA-2013-BLG-220 was conducted in 2015 using Keck's NIRC2 Adaptive Optics system with the wide camera, and again in 2019 using the narrow camera from the same instrument. Before analysing the 2015 data we first reprocessed the cube of JHK data obtained as part of the VVV survey with the ESO VISTA 4m telescope centered on the target following
the same procedure as in \cite{Beaulieu2018}. This  gave us a reference catalog both for photometry and astrometry. 

In 2015, the images taken with the wide camera had a pixel scale of 0.04 arcsec per pixel. The \textit{$K_s$} filter (K-short, hereafter \textit{K} band) was used which resulted in 22 images of 30 seconds each, taken at 5 dithered positions. Following a standard procedure outlined in \cite{Beaulieu2016} and \cite{Batista2014}, we subtracted the dark-current and flat-field before
performing astrometry and then coadded the frames.  The resulting stacked image had a full-width-half-maximum (FWHM) of 100 mas. Using MOA images at high amplification we refined the RA, DEC position of the source to be (18:03:56.50~-29:32:41.0, J2000.0). The source and lens were not resolved in 2015.

We ran the SExtractor program \citep{Sextractor96} to measure fluxes of the different sources in the field  and calibrate them by cross-identifying the 
stars with the VVV catalog.  We estimated that the error of the zero point of our calibration with 40 stars is 1.1 \%. Thus, we measured $K_{\rm source+blend} = 16.75 \pm 0.02$.

On 27th May 2019, further high angular resolution follow-up observations of MOA-2013-BLG-220 were carried out using Keck's NIRC2 Adaptive Optics system. The images were taken using the narrow camera with a plate scale of 0.01 arcsec per pixel. We obtained 21 usable \textit{K} band images with point spread function (PSF) FWHM of $\sim 50-60$ mas. The exposure times of these images are 30 seconds, with a dither of $\sim 2$ arcsec. The images revealed that the source and lens already have a visible separation. From \cite{Yee2014} the lens and source have a relative proper motion of $\sim12$ mas per year. According to this, the predicted separation between the source and lens in May 2019 should be $\sim75$ mas. Therefore, we expect to be able to measure the magnitude for each star individually.

Similarly to the 2015 data, this analysis included dark-current, flat-field and sky correction. The narrow images from 2019 were calibrated to the wide images from 2015 which helped cross-identify isolated stars. The frames were then median stacked using SWARP \citep{Bertin2010}. We then continued the following analysis using the 2019 data set.

\begin{figure*}
	\centering
	\includegraphics[width=16cm]{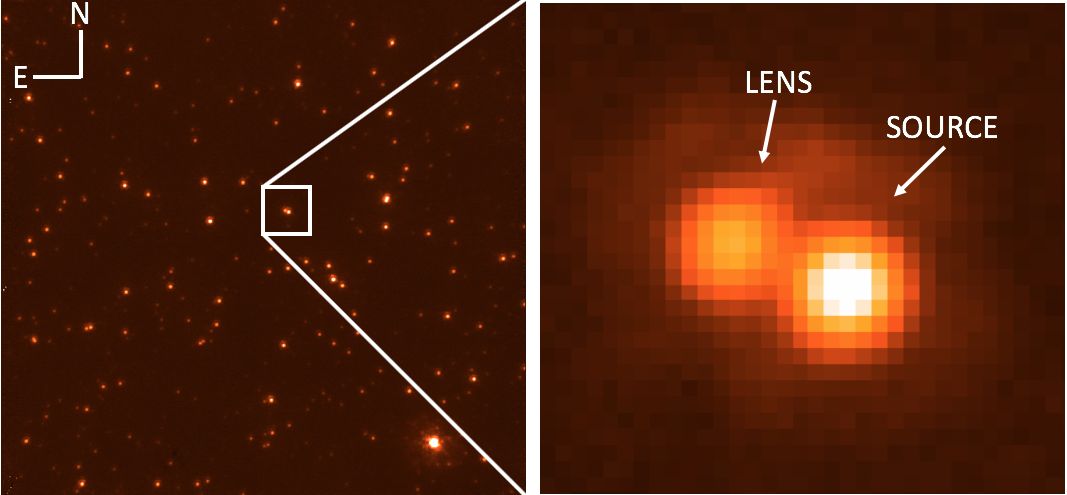}
	\caption{Keck image of MOA-2013-BLG-220 in \textit{K} band with the NIRC2 AO system, taken on 27th May 2019. On the left, the image is $10'' \times 10''$ and on the right the panel is a zoom of approximately $ 1'' \times 1''$. The right panel also indicates the position of the lens and source.} 
	\label{220_v5}
\end{figure*}

 We identified the brighter star as the source (see Figure \ref{220_v5}). This was confirmed by the measured K-band magnitude of the source, which matched the predictions based on the MOA-red and CTIO-$V$ source magnitudes (see Section \ref{sec-lc_model}).
 
\section{Relative source lens proper motion and flux ratio}

 The source and lens were unresolved within the 100 mas resolution of the 2015 Keck data. The combined flux was measured, but the possible contamination by additional blended sources could not be assessed. 
	In 2019 we obtained higher angular resolution observations with Keck's narrow camera (50-60 mas resolution). With the 2019 data we resolved source and lens, and measured their flux ratio, amplitude and direction of their relative proper motion.
	We expected that the separation between source and lens would be of the order of  $\sim 1 $ FWHM, so using PSF fitting
	photometry is the way forward.  One approach could have been to do a double star fit using tools 
	such as DAOPHOT \citep{Stetson1987}. However with adaptive optics observations, we have a complex 
	PSF with large wings and that is spatially varying in the field. Our objective is to measure the flux ratio and separation of the source and lens. Therefore instead of using DAOPHOT, we decide to adopt a new approach that will also be applied to systems where the source and lens are not resolved, such as \cite{Bhattacharya2018}.
	Our aim is to provide accurate error estimates and covariances between the main parameters of interest: orientation and amplitude of the relative source-lens proper motion and flux ratio. It should be noted that since the two objects are
	separated by less than 100 mas, they have the same PSF shape.

The first step is to make a numerical estimate of the PSF. The PSF is reconstructed on
a grid by stacking the brightest stars in the neighbourhood of the object of interest. 
The accurate position of each PSF is estimated
by iterative Gaussian weighted centering. The PSF's are then interpolated and recentered
in a common reference frame. Finally, stack the PSF's and take a median value.
The reconstruction of a system involving two close stars with the same PSF
involves six independent parameters. These are the position of the two stars $( {X_A,X_B})$
(four parameters) and the flux of the two stars, $A$ and $B$. However, the number of effective parameters
can be greatly reduced if we use some basic constraints on the parameters. If the photocenter of the system is a known quantity we introduce two constraints
on the system, and if the total flux is known, we introduce an additional constraint.
As a consequence, we are left with the reconstruction of only three independent parameters.

It is convenient to work in the reference frame of the system's photocenter (G).
In this frame a line passing through the photocenter of each star also intercepts the origin of the coordinate system. This system is defined by three parameters: the orientation of the line joining the points, $\theta$, the flux ratio, $\xi=\frac{A}{A+B}$, and the angular separation of each point from the photocenter, ($S_A$;$S_B$). It is more convenient to use the the total separation $S=S_A+S_B$ (which is the separation between the two stars) and the separation ratio $\xi_S=\frac{S_A}{S}=1-\xi$ rather than $S_A$ and $S_B$.
These three parameters can be related to the other forementioned parameters through the following equations:

\begin{gather*}
 X_A= X_G-\xi_S S {\bf U_r} \\
X_B= X_G+(1-\xi_S) S {\bf U_r} \\
A=\phi_0 (1-\xi_S) \\
B=\phi_0 \xi_S
\end{gather*}

Here we define the vector $  U_r = \left( \cos(\theta),\sin(\theta) \right)$ and $\phi_0$ as the total flux of the two stars. In practice it is
convenient to estimate the PSF in a reference frame where its photocenter is at the origin. By doing so,
the PSF coordinates are also the coordinates of its
photocenter.  In this method, we define the reference position to be the photocenter of the combined sources. This position is far (about 4 pixels) from the peak of the PSF of the brighter star, so using the star position as a first guess for the reference position leads to a very inefficient search, and typically a failure of the method. Instead, we take as our first estimate of the photocenter position the location of the zeroes of the first two moments of the light distribution. This position is refined using a nonlinear optimization, minimizing $\chi^2$ over a grid of positions. The uncertainty in the position thus obtained is $\approx$ 0.2 pixels.

Since we have only the three independent parameters  $(\theta,\xi_S,S)$ it is possible to
implement a simple grid search for the optimal $\chi^2$. This basic method
could be applied directly to the data, but in practice one has to consider that the estimates
of the total flux and photocenter from the data are quite imperfect and noisy, and as a consequence may affect
the quality of the reconstruction. The aperture evaluation, which evaluates a quantity by using the pixels within an aperture radius, may be particularly inefficient and sub-optimal if the PSF has large wings, which is the case here.

As a consequence, we evaluate the total flux and photocenter in an aperture radius. Using this as a first guess, we conduct a non-linear optimization of these three parameters to minimize the $\chi^2$ for each estimate of the other three parameters $(\theta,\xi,S)$. Since the initial estimation of the total flux and photocenter is by construction close to the real value, the non-linear refinement procedure always converges quickly with only a small change of the initial guess. This eliminates degeneracy in the refinement procedure, and so, we are only left with a grid search for the best $\chi^2$ in the $(\theta,\xi,S)$ space. Finally, we take the best
value of the  $(\theta,\xi,S)$ and their associated refined total flux and photocenter position,
and perform a small non-linear optimization of all six parameters in the vicinity of the $\chi^2$ minimum, to
obtain the final parameter values.  Note that for this particular case there is no ambiguity or degeneracy in the solution since the 
two components are clearly resolved in the image. We also checked the local grid around the best parameters and found no 
indication of a correlation between the parameters. Thus we find the source-lens angular separation $S=77.63 \pm 0.24$ mas, flux ratio $\xi = 0.3367 \pm 0.0025$ and angle
$\theta = 2.7712 \pm 0.0035$ radians. 

\begin{figure}
   	\centering
   	\includegraphics[width=8.5cm]{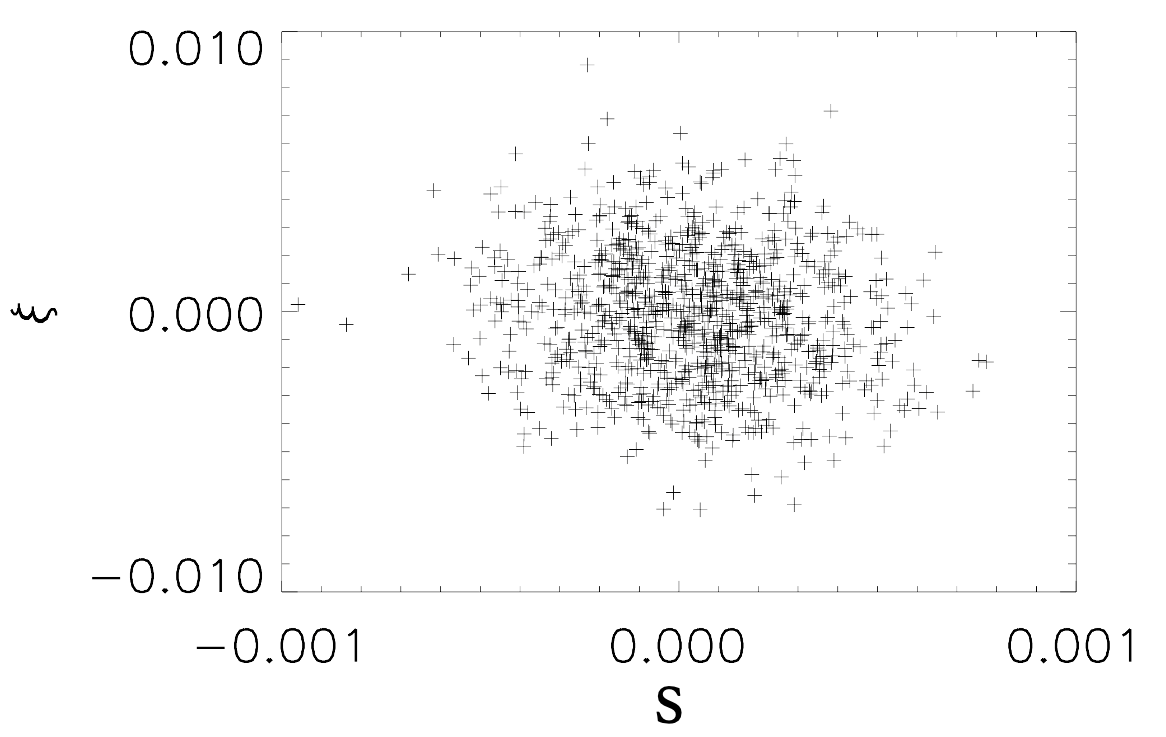}
   	\caption{Estimating errors on the parameters flux ratio $\xi$ and angular separation $S$ (in arcsec) by using numerical simulations. A total of 1000 simulations were used in this plot. } 
   	\label{Sokolov}
   \end{figure}

The final parameter estimates depend on the distribution of observational parameters, which are highly correlated. To estimate the uncertainties we take a Monte Carlo approach, sampling from the Poisson distribution for each variable and re-running our fit procedure 1000 times to simulate the likely range of derived values. This yields robust estimates of the uncertainties in separation S, flux ratio $\xi$, and orientation angle $\theta$. Because the uncertainties on the input parameters are small and there are only a small number of variables with a single well-defined global minimum, it is not necessary to re-run the entire grid search for each iteration; we apply the Monte Carlo procedure only to the final non-linear optimization. Thus we find the standard deviation for the separation $\sigma_S=0.024$ and for the flux ratio $\sigma_{\xi}=0.0025$. For a general view of the simulations results in the separation versus flux ratio plane, see Figure ~\ref{Sokolov}.

   \begin{figure}
	\centering
	\includegraphics[width=8.5cm]{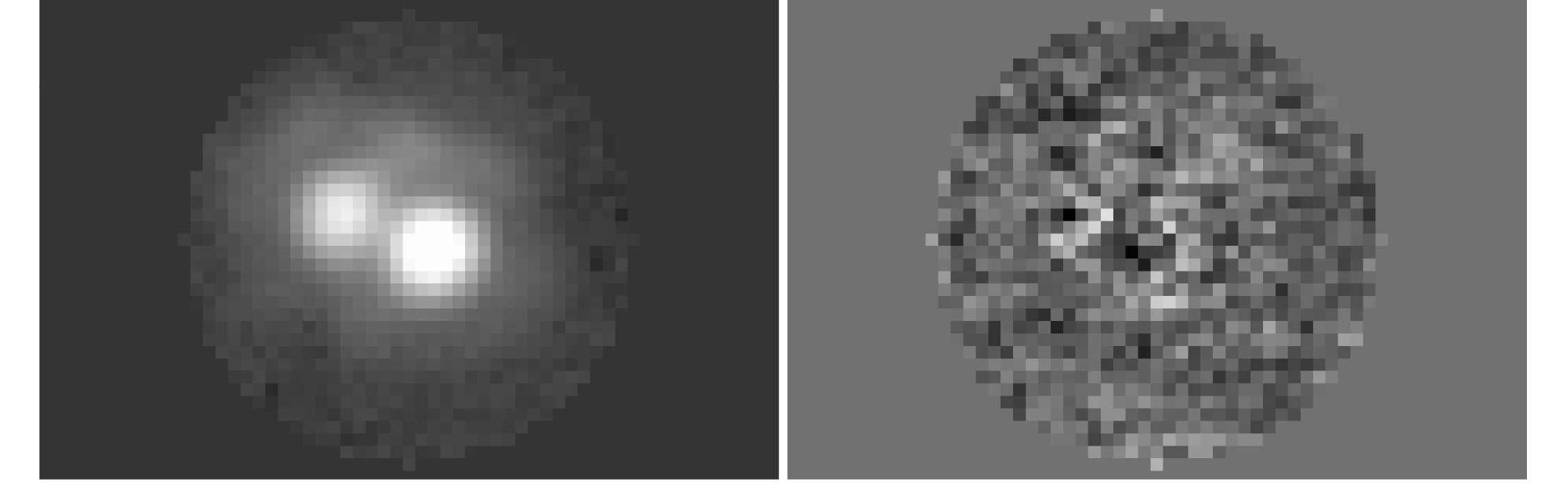}
	\caption{Fit of source and lens (left) and the residuals after subtracting the model (right).} 
	\label{fit220}
\end{figure}

As a sanity check, we redid the above analysis using a different approach. Instead of stacking all our frames and having a single image to work with, we separated our frames into three groups which were then stacked using SWARP. This method was explored in case the previous method had significantly underestimated uncertainties that are dominated by systematic errors. Therefore, we find three parameter values for each stack (see Table \ref{tablex}) taking the average value as our result and the scatter as our error estimate.

\begin{deluxetable}{ c | c c c }[h!]
	\tablecaption{ Parameter results for the three sets of stacked images.  \label{tablex} }
	\tablewidth{0pt}
	\tablehead{
		           & \colhead{Stack 1}  & \colhead{Stack 2} & \colhead{Stack 3} 
	}  
	\startdata
	S (mas) & 76.791 & 77.641 & 76.964 \\  
	$\theta$ (radians) & 2.7750 & 2.7674 & 2.7616 \\ 
	$\xi$ & 0.3290 & 0.3344 & 0.3340 \\
	\enddata
\end{deluxetable}	

Therefore, we find $S = 77.13 \pm 0.45$ mas, $\theta = 2.7680 \pm 0.0067$ radians and $\xi = 0.3323 \pm 0.0030$, which are in agreement with our values found using the first method. Since the values and associated errors found using the second method are more conservative, we continue with these.

 Knowing the flux ratio $\xi = 0.3323 \pm 0.0030$  and the calibrated source and lens flux in $K$ band from the 2015 observation $K_{\rm source+blend} = 16.75 \pm 0.02$,  we derive the $K$ magnitude of the source and the lens :
\begin{gather*}
K_{\rm Keck,lens} = 17.92 \pm 0.02\\
K_{\rm Keck,source} = 17.20 \pm 0.02
\end{gather*}

The follow up observations were taken 6 years after the initial observation in 2013, therefore we can deduce that the heliocentric relative proper motion of source and lens is 
$\mubold_{\rm rel, Hel} = (9.74\pm 0.09,  -8.05\pm 0.09)$, in Galactic coordinates (longitude and latitude, or $l, b$).
The magnitude of the relative proper motion vector is $\mu_{\rm rel,Hel} = 12.54 \pm 0.13$ $\rm mas/ yr$. 

The light curve model uses a geocentric reference frame however, so our relative proper motion magnitude 
needs to be converted using the following relation \citep{Dong2009a}:
\begin{equation}
\mubold_{\rm rel, Hel} = \mubold_{\rm rel,Geo} + \frac{\boldsymbol{\nu}_{\oplus}\pi_{rel}}{AU}
    \label{eq-murelH}
    \end{equation}
where, $\boldsymbol{\nu}_{\oplus}$ is the Earth's projected velocity relative to the Sun at the time of peak magnification. For MOA-2013-BLG-220, this is $\boldsymbol{\nu}_{\oplus}=(3.2,6.6)\ \rm kms^{-1}$ \citep{Yee2014}. Since the distance to the lens and the source is great, the difference between $\mubold_{\rm rel}$ in geocentric and heliocentric coordinates is insignificant, as we calculate a $\mubold_{\rm rel,Geo} = 12.62 \pm 0.11$ $\rm mas/ yr$.

\section{Planetary System Parameters}
\label{planet}

   \begin{figure*}
   	\centering
   	\includegraphics[width=12cm]{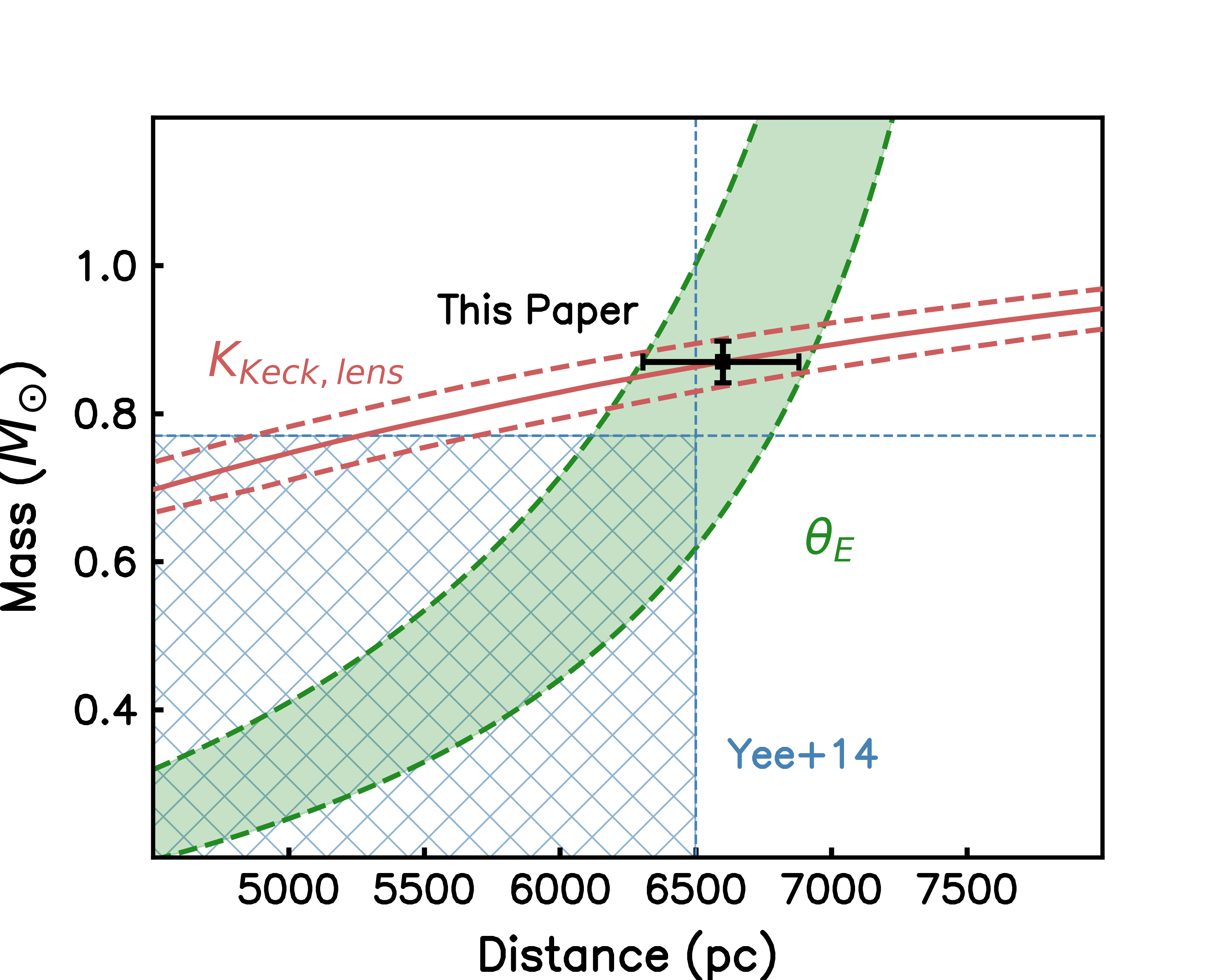}
   	\caption{Mass-Distance relations for MOA-2013-BLG-220, 
   		obtained from the Einstein radius constraint (green) and K-band flux constraints from Keck observations taken this year (red). \cite{Yee2014} upper limit estimates are also indicated in blue.} 
   	\label{md220}
   \end{figure*}
   
One way to determine the lens mass and distance to the planetary system is through the intersection of the mass-distance relations. This can be established by combining the lens magnitude measurement with empirical mass-luminosity relations. The first mass-distance constraint is derived from the  Einstein ring radius, $\theta_E$, which equals to $0.457 \pm 0.005\ \rm mas$ in this case. Because both stars are resolved, $\theta_E$ is very tightly constrained. The error seen on Figure \ref{md220} for this constraint is primarily due to the uncertainty on the distance to the source, which for this event is $D_S = 8.19 \pm 0.76$ kpc, as found from the MCMC simulation. The first constraint is shown in equation \ref{einstein}.

\begin{equation}
\label{einstein}
M_L = \frac{\theta_E^2}{\kappa \pi_{ rel}},\ \ \ \ \
\pi_{ \rm rel} = AU(\frac{1}{D_L} - \frac{1}{D_S})
\end{equation}

Where $M_L$ is the lens mass, $D_L$ is the lens distance and $\kappa = 8.144\ \rm mas\ M_\odot^{-1}$.

An additional constraint on the mass-distance relation can be implemented through the combination of isochrones with the measured NIR magnitude for the lens from Keck. This constraint is expressed as: 
\begin{multline}
 m_{L} = 10 + 5\log(D_{L}/1{\rm kpc}) +A_{K,L}+ \\ 
 M_{\rm isochrone}(\rm \lambda, M_{L}, age, [Fe/H])
 \label{eq-mass_dist}
\end{multline}

Where $M_{\rm isochrone}$ is the predicted absolute magnitude for a given mass ($M_L$) of the lens, age and metallicity. The interstellar extinction in \textit{K}-band along the lens' line of sight is given by $A_{K,L}$. The extinction is determined from the \textit{VVV Extinction Calculator} \citep{extinction-vvv}
and we obtain $A_{K,L} = 0.17 \pm 0.02$. The isochrones used for this constraint are the 6.4\ Gyr population from \cite{Girardi2002}, and the error on lens mass and distance take into account age uncertainty. Plotting these constraints on Figure \ref{md220} we can determine the mass and distance of the lens from where both intercept. The isochrone constraint is shown in red, where the dashed lines indicate the error on the measured lens magnitude. The $\theta_E$ constraint is shown in green. 

\cite{Yee2014} provides constraints on the lens mass and distance based on an estimate of an upper limit on the lens brightness, 
which is shown in blue in Figure~\ref{md220}. This limit stems from the source having an absolute magnitude of $\sim 3.4$, and the lens (being at a closer distance) being at least $\sim 1$ magnitude dimmer, otherwise it would have an affect on the blended light. Using \cite{Yee2014}'s measurement of $\theta_E$ and assuming that the lens is in or near the bulge, they derive $M \simeq 1.7M_{\odot}(kpc/D_{S}-D_{L})$. Therefore, they get a lens that is at least 1.7 kpc in front of the source, providing an estimate for lens distance, $D_L <6.5$ kpc, and mass, $M<0.77\ M_\odot$.
However, \cite{Yee2014} did not include the contribution of unresolved stellar flux to the background in seeing-limited images. This is a weakness of attempts to constrain the blend fraction from low angular resolution observations.

\begin{figure*}	
	\centering
	\includegraphics[width=17cm]{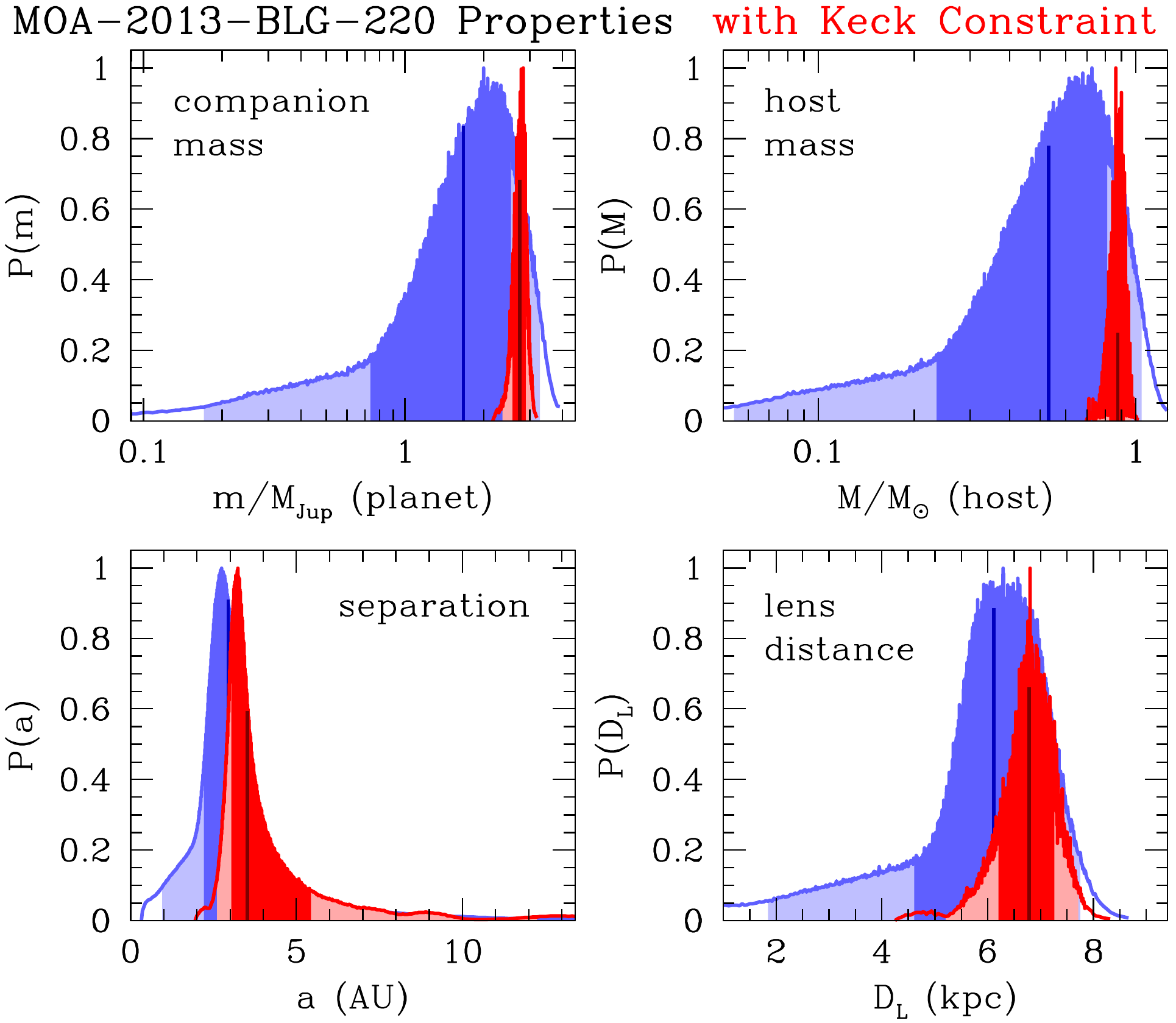}
	\caption{Distributions for the Bayesian posterior probability for the planetary and host star mass, their separation and the distance to the lens. These are shown with only light curve data (blue) and with additional constraints from Keck (red). The central darker shades represent the 68.3$\%$ of the distributions, and the lighter shades the 95.4$\%$ of the distributions. The black vertical line indicates the median of the probability distribution for each parameter}
	\label{fig-model}
\end{figure*}

A more precise measurement of the lens properties can be determined by combining our MCMC light curve
model distribution with our constraint on the lens-source relative proper motion of
$(\mu_{\rm rel, H,l}, \mu_{\rm rel, H,b}) = (9.74\pm 0.09,  -8.05\pm 0.09)$, and a Bayesian prior from a Galactic
model \citep{Bennett2014}. The results of this calculation are shown as the red histograms in Figure~\ref{fig-model}.
The blue histograms in this figure show the predicted probability distribution of the planet and host masses,
projected separation and lens distance without the constraints from our Keck observations. The calculations
used to produce the blue histograms make the assumption that 
the probability of hosting this planet is independent of the host mass. This is the simplest assumption to make,
but we do not have much justification for it.

Our calculations with the Keck constraints (the red histograms in Figure~\ref{fig-model}) indicate a lens
(and host) star mass of $M_{L} = 0.88 \pm 0.05\ M_{\odot}$, orbited by a planet of mass
$M_{p} = 2.74 \pm 0.17\ M_{J}$, at a distance of $D_{L} = 6.72 \pm 0.59$ kpc. The planet-star mass 
ratio is $q = (3.00 \pm 0.03) \times 10^{-3}$ and the projected star-planet separation is
$ r_\perp = sD_{L}\theta_E = 3.03 \pm 0.27\,$AU. So, this system consists of a super-Jupiter mass 
gas giant planet orbiting a late G-dwarf star at a separation beyond the snow line.

\section{Discussion and Conclusion}
\label{final}

We have detected the planetary host star for microlensing event MOA-2013-BLG-220, and determined that the system comprises of a gas giant orbiting a solar-type star. Using high angular resolution follow-up observations from Keck, 6 years after the event was detected, we have resolved the source and the lens to a separation of $\sim 78$ mas in the \textit{K} band, enabling us to accurately measure their flux.

This event was of particular interest for a follow up observation because of its high relative proper motion of $\sim 12.6$ mas/year. High relative proper motions can be produced by either the lens and/or source moving fast relative to their population, or other nearby lenses. In this case, the source star was observed 1298 tmes between 2001 and 2009 by OGLE-III, and therefore its proper motion was measured and found to be $\mu_{\rm base} = (\mu_{l}, \mu_{b})=(-5.6, +1.9)$ mas/yr, indicating that it is moving directly opposite to the direction of Galactic rotation. In the discovery paper \cite{Yee2014} argue that the combination of a source star in retrograde motion, combined with a lens in the Galactic disk, would produce the observed relative high proper motion. 

From the results in this paper we find that the lens is more likely to be in the bulge ($D_{L} \sim 7$ kpc). A bulge lens does not rule out a high relative proper motion however. The proper motions of stars in the bulge have high velocity dispersions that could indeed account for this event's high proper motion \citep{Portail2017, Portail2015}. At the galactic latitude of this event, $(l, b) = (+1.5, -3.76)$ the mean velocity dispersion in the bulge is $\sim 100$ km/s, as shown in Figure 15 from \cite{Portail2017}. 

Furthermore, at low galactic latitudes there is a mix of disk, bar, and bulge populations along sightlines toward the bulge.  Bar stars in particular are known to be characterised by bulk streaming motions with high velocities (e.g. \cite{Valenti2018}, \cite{Sanders2019}).

Using these constraints and additional photometric light curve data that were excluded by \cite{Yee2014}, we improved upon previous modeling and refined system parameters. Our results indicate that the system is composed of a $M_{L} = 0.88 \pm 0.05\ M_{\odot}$ star orbited by a $M_{p} = 2.74 \pm 0.17\ M_{J}$ planet at a projected separation of $3.03 \pm 0.27$ AU. This of course is just the instantaneous projected separation; the eccentricity and orbital semi-major axes are unknown. However, true orbital radii must be equal to or larger than the projected separation, which puts this gas giant beyond the snow line for a late G-type star. Figure \ref{fig-model} demonstrates these results and in particular, emphasizes the importance of follow up observations and the tight constraints that they place on the physical parameters of the system.

One of the most interesting features of Figure \ref{fig-model} is that the host mass is much more massive than the
 median prediction made from the Bayesian analysis, which assumes that stars of all masses are equally likely to host planets of a given mass ratio (i.e., a uniform prior on the distribution of mass ratios with planetary host mass was assumed). This assumption is often used to estimate the host mass for many planets found via microlensing (e.g. \cite{Beaulieu2006}, \cite{Dong2009a}), though it has not yet been extensively tested. High-resolution observations however, can provide the opportunity for this assumption to be tested.

In the case of MOA-2013-BLG-220 and MOA-2007-BLG-400 (\cite{Dong2009b}, Bhattacharya et al., (in prep)), the measured host mass is much larger than the median Bayesian estimate. Both are bulge stars more massive than 90$\%$ of stars in the predicted mass distribution function, with a mass ratio of $q \sim 2 \times 10^{-3}$. Other events, with both larger \citep{Udalski2005,Dong2009b,Bennett2019}  and smaller \citep{bennett15,Batista2015}) mass ratios, have host much closer to the median Bayesian prediction. It is possible that the mass ratio distribution varies with stellar mass, but clearly more mass measurements for microlensing planets are required in order to explore this possibility.

Analyses of radial velocity data (e.g. \cite{Johnson2007, Johnson2010}) revealed a positive correlation between the frequency of giant planets and stellar host mass. This was also observed in a direct imaging survey by \citet{Nielson2019}, where more massive stars were found to be more likely to host massive planets at wider orbits compared to smaller stars. Microlensing events MOA-2013-BLG-220 and MOA-2007-BLG-400 also agree with this. Radial velocity surveys (e.g. \citet{Gonzalez1997, Fischer2005}), have also indicated that giant planet occurrence increases for more metal rich stars.  In the future this can be tested further by determining the host star's metallicity from spectra, using the Keck/Osiris and VLT/MUSE instruments. For planets found via microlensing, this will only be possible where the source and lens star can be resolved, such as MOA-2013-BLG-220Lb and OGLE-2005-BLG-071Lb.
 
This analysis demonstrates the benefits of high angular resolution follow-up observations several years after the event has occurred, because the lens and source can be resolved. This allows for accurate measurements of the lens magnitude, which when combined with other constraints (such as $\theta_E$), can help determine the physical parameters of the system. Having precise mass measurements of both lens and planetary companion is key to understanding planet formation. \\

		\acknowledgments
	This work was supported by the University of Tasmania through the UTAS Foundation and the endowed Warren Chair in Astronomy and the ANR COLD-WORLDS (ANR-18-CE31-0002). This research was also supported in part by the Australian Government through the Australian Research Council Discovery Program (project number 200101909) grant awarded to Cole and Beaulieu. The Keck Telescope observations and analysis were supported by a NASA Keck PI Data Award, administered by the NASA Exoplanet Science Institute. Data presented herein were obtained at the W. M. Keck Observatory from telescope time allocated to the National Aeronautics and Space Administration through the agency’s scientific partner- ship with the California Institute of Technology and the University of California. The Observatory was made possible by the generous financial support of the W. M. Keck Foundation. DPB and AB were also supported by NASA through grant NASA-80NSSC18K0274. 
	This work has made use of data from the European Space Agency (ESA) mission
{\it Gaia} (\url{https://www.cosmos.esa.int/gaia}), processed by the {\it Gaia}
Data Processing and Analysis Consortium (DPAC,
\url{https://www.cosmos.esa.int/web/gaia/dpac/consortium}). Funding for the DPAC
has been provided by national institutions, in particular the institutions
participating in the {\it Gaia} Multilateral Agreement.

	\bibliographystyle{yahapj}
	\bibliography{220PAPER}
\end{document}